\documentclass[iop]{emulateapj}
\usepackage{epsfig}
\usepackage{graphicx}
\usepackage{epstopdf}
\usepackage{morefloats}

\shorttitle{$\delta$~Scorpii's Circumstellar Disk}
\shortauthors{Jones et al.}

\begin{document}

\title{Using Photometry to Probe the Circumstellar Environment of $\delta$~Scorpii}
\author{C. E. Jones\altaffilmark{1}, Paul Wiegert\altaffilmark{1}, C. Tycner\altaffilmark{2}, G. W. Henry\altaffilmark{3}, R. P. Cyr \altaffilmark{1}, R. J. Halonen \altaffilmark{1}, M. W. Muterspaugh\altaffilmark{3,}\altaffilmark{4}}
\altaffiltext{1}{Department of Physics and Astronomy, Western University, London, ON Canada N6A 3K7}
\altaffiltext{2}{Department of Physics, Central Michigan University, Mt. Pleasant, MI 48859 USA}
\altaffiltext{3}{Center of Excellence in Information Systems, Tennessee State University, 3500 John A. Merritt Blvd., Box No 9501, Nashville, TN 37209, USA}
\altaffiltext{4}{Department of Mathematical Sciences, College of Engineering, Tennessee State University, Boswell Science Hall, Nashville, TN 37209, USA}

\begin{abstract}

We acquired Johnson $BV$ photometry of the binary Be disk system 
$\delta$~Scorpii during its 2009, 2010, 2011, and 2012 observing
seasons and used it to probe the innermost regions of the disk.
We found that several disk building events have occurred during this time, resulting in an overall brightening in the $V$-band and reddening of the system. In addition to these long-term trends, we found cyclical variability in each observing season on timescales between 60 and 100 days. We were able to reproduce the changes in the magnitude and colour of $\delta$ Sco using our theoretical models and found that variable mass-loss rates in the range $2.5 - 7.0 \times 10^{-9}$
$M_{\sun}/$yr over $\sim$ 35 days can reproduce the observed increase in brightness. 
\end{abstract}


\section{Introduction}

The classical Be (B-emission) stars are well known for their characteristic spectral emission lines in the Balmer series, the defining property of the group, and infrared excess due to radiative processes occurring within the disk-like distribution of circumstellar material. Be stars are also characterized by rapid rotation, of typically several hundred km s${^{-1}}$, and this property certainly plays a role in the release of material from the stellar surface to form a disk. Their rotation rates are difficult to determine because of the effects of gravity darkening and, as a result, the precise values remain a contentious issue \citep{tow04,cra05}. Despite these unknowns, this group of stars offers a valuable test bed to study disk physics and the effects of rapid rotation on stellar evolution. 

In addition to rapid rotation, emission lines, and infrared excess, there are other commonly observed features among this group of stars, such as polarized continuous light due to Thomson scattering of stellar radiation and variability on a range of time-scales from minutes to decades \citep{por03}. The longer time scales are associated with significant disk building and disk dissipation events. Short term variations on typical scales of 0.5 to 2 days are often associated with stellar pulsation \citep{riv03}.

There have been a variety of models proposed to explain the formation of these disks and associated characteristics beginning with \citet{str31} who suggested that Be star disks are formed exclusively by the rapid rotation of the central star with the variety of spectral line shapes due to viewing angle. However, this view may be too simplistic, since it is generally believed that Be stars rotate below critical so there must be another mechanism(s) to lift material off the stellar surface. 

Though there have been many different models proposed since Struve's time to explain these systems, the viscous disk model has been most promising \citep{car11}. Basically, this model operates similarly to the standard $\alpha$-disk theory for accretion disks of inward flowing material except, in this case, the gas is outward flowing. The viscous model was first applied to Be stars by \citet{lee91}, later investigated by \citet{oka01}, and more recently is being used to explain Be disk systems in exceptional detail \citep{car06sco,car12a}. This model naturally predicts near-Keplerian disk rotation, a property that is consistently implied through observations \citep{mei07,oud11,whe12}. Viscous disks may also become unstable to density perturbations allowing the formation of under- and over-dense volumes of gas that could give rise to the characteristic variability of doubly peaked emission lines that are often observed. Viscous disk models have also successfully predicted the observed IR excess (see \citet{car09}). 

The actual mechanism for launching material from the rapidly rotating star remains unknown, and often the value of the mass loss rate is a free parameter in the modeling process (see e.g., \citet{car06sco}). $V$-band photometry has been shown to probe the innermost regions of Be disks \citep{car11} and so will show the quickest reaction to photospheric activity \citep{car12b}. Therefore, $V$-band photometry combined with modeling may be used to place bounds on the mass-loss rates and contribute to a better understanding of the launching mechanism(s). 

Interestingly, $\delta$~Sco (HD143275, HR5953) was originally designated as an $uvby\beta$ standard star of spectral type B0.3IV \citep{per87} but later \citet{cot93} observed emission in the wings of H$\alpha$ providing evidence that $\delta$ Sco had become a Be star. In that same year, \citet{bed93} confirmed spectroscopic binarity and reported an elliptical orbit with a 10.5 year orbit. A significant brightening occurred in 2000 \citep{ote01} and since that time, the highly variable early-type Be star $\delta$~Sco has been a focus of much attention. It is interesting to speculate that the periastron passage in 2000 contributed to the observed brightening. The secondary companion has a very eccentric orbit with a period of about 11 years (for recent calculations see \citet{tan09,tyc11,mei11}) and the periastron passage of its companion in 2011 offered an opportunity to study this system. We collected $V$- and $B$-band photometry for the 2009, 2010, 2011 and 2012 observing seasons with the T3 0.4m automatic photoelectric telescope (APT) operated by Tennessee State University (TSU) and located at Fairborn Observatory in southern Arizona. 

In this work we use our $V$ and $B$ photometry to probe the innermost regions of the disk of $\delta$~Sco in order to determine the mass loss rate and investigate the variability of these outbursts for this system. Our observations and results are presented in Section~\ref{ob} and Section~\ref{results}, respectively, and a discussion and summary is provided in Section~\ref{sum}.

\section{Observations}
\label{ob}
The T3 0.4~m automatic photoelectric telescope (APT) at Fairborn Observatory 
in southern Arizona acquired several hundred observations of $\delta$~Sco 
during the 2009, 2010, 2011, and 2012 observing seasons.  T3 is one of eight 
automatic telescopes operated by Tennessee State University at Fairborn for 
automated photometry, spectroscopy, and imaging \citep{Henry95, Henry99, 
ehf03, ew07}.  T3 is equipped with a precision photometer that employs an 
EMI~9924B photomultiplier tube (PMT) for the successive measurements of 
photon count rates through Johnson $B$ and $V$ filters.  To maximize the 
stability of the photometer and the precision of the data, the PMT, voltage 
divider, pre-amplifier electronics, and photometric filters are all mounted 
within the temperature- and humidity-controlled body of the photometer.  
The precision of a single observation on a good night is usually in the range 
$\sim~0.003 - 0.005$ mag \citep[e.g.,][Table 20]{Johnson11}, depending 
primarily on the brightness of the target and the airmass of the 
observation.  Year-to-year brightness means are stable to 0.0001 - 0.0002
mag over decadal time scales \citep{Henry99}.

We programmed the APT to make one or two observations of $\delta$~Sco each 
clear night in the following sequence, termed a group observation: K,S,C,V,C,V,C,V,C,S,K, where K is the check star (HD~144217, $V = 2.56$, 
$B-V = -0.06$, B0.5~V), C is the comparison star (HD~144470, $V = 3.93$, 
$B-V = -0.05$, B1~V), V is the program star $\delta$~Sco ($V=2.29$, 
$B-V = -0.12$, B0.2~IVe), and S is a sky reading taken near the center of 
the group.  Because all three of these stars are bright, we used a 3.8-mag 
neutral density filter in addition to the Johnson filters to attenuate the 
count rates and therefore avoid saturation of the PMT and minimize the deadtime.  Each group 
observation is reduced to form three bracketed $V-C$ and two unbracketed 
$K-C$ differential magnitudes, which are averaged together to create group 
means for both $B$ and $V$ bands.  Group-mean differential magnitudes with 
internal standard deviations greater than 0.01 mag were discarded to 
eliminate observations taken under non-photometric conditions.  The surviving 
group means were corrected for differential extinction with nightly 
extinction coefficients, transformed to the Johnson system with yearly-mean 
transformation coefficients, and treated as single observations thereafter.  

T3 acquired a total of 511 group observations over the four observing 
seasons of our campaign.  The observations that survived the cloud-filtering 
process are plotted in the four panels of Figure~\ref{photo_all} and listed in Table~\ref{photo_ob}. All four panels of Figure~\ref{photo_all} are plotted using the same vertical and horizontal scales to allow direct comparison 
of the brightness variations in the $V$ and $B$ filters, in the $B-V$ color 
index, and any brightness changes in the $K-C$ differential magnitudes during the four seasons of our observational campaign. The $K-C$ observations scatter around their grand mean with a standard 
deviation of 0.0089 mag.  This is somewhat larger than the typical precision 
of $\sim~0.004$ mag referenced above, primarily because $\delta$~Sco lies at 
a declination of $-20\arcdeg$ and so is observed through a larger than 
usual air mass. The $K-C$ observations in the bottom panel of Figure~\ref{photo_all} are quite flat compared to the $V-C$ observations plotted in the top three panels, indicating that the variability in the $V-C$ observations are intrinsic to $\delta$~Sco.  We do suspect slight variability in the
comparison star at the very end of our time series (see Section 3 below). The arrow in the top panel marks the time of the latest periastron passage on UT 2011 ${\rm July}~6 \pm 2$ days, as
predicted by \citet{tyc11}.
  
\begin{deluxetable}{ccccc}
\tablecolumns{5}
\tablewidth{0pt}
\tablecaption{Photometric Observations of $\delta$~Sco \label{photo_ob}.}
\tablehead{
\colhead{Reduced Julian Date} & \colhead{Var $B$} & \colhead{Var $V$}
& \colhead{Chk $B$} & \colhead{Chk $V$} \\
\colhead{} & \colhead{(mag)} & \colhead{(mag)} & \colhead{(mag)}
& \colhead{(mag)} 
}
\startdata
54,890.0405 & $-$1.933 & $-$1.951 & $-$1.475 & $-$1.452 \\
54,891.9763 & $-$1.930 & $-$1.939 & $-$1.480 & $-$1.444 \\
54,892.0043 & $-$1.922 & $-$1.942 & $-$1.475 & $-$1.447 \\
54,898.9435 & $-$1.913 & $-$1.907 & $-$1.492 & $-$1.440 \\
54,901.9344 & $-$1.896 & $-$1.884 & $-$1.487 & $-$1.453 \\
54,901.9752 & $-$1.892 & $-$1.881 & $-$1.475 & $-$1.445 \\
54,902.9295 & $-$1.880 & $-$1.885 & $-$1.480 & $-$1.447 \\
\enddata

\tablecomments{Table~\ref{photo_ob} is presented in its entirety in
the electronic edition of the Astrophysical Journal.  A portion
is shown here for guidance regarding the form and content. }
\end{deluxetable}

Figure~\ref{photo_09_10} and Figure~\ref{photo_11_12} plot the individual seasons of $\Delta V$ magnitudes and 
$\Delta(B-V)$ color indices to facilitate their comparison.  The vertical axis 
scale of the individual panels in those two figures are no longer the same 
as they were in Figure~\ref{photo_all} but are chosen individually to best display the 
individual light curves.

\section{Results}
\label{results}

\begin{figure}
\epsscale{0.75}
\plotone{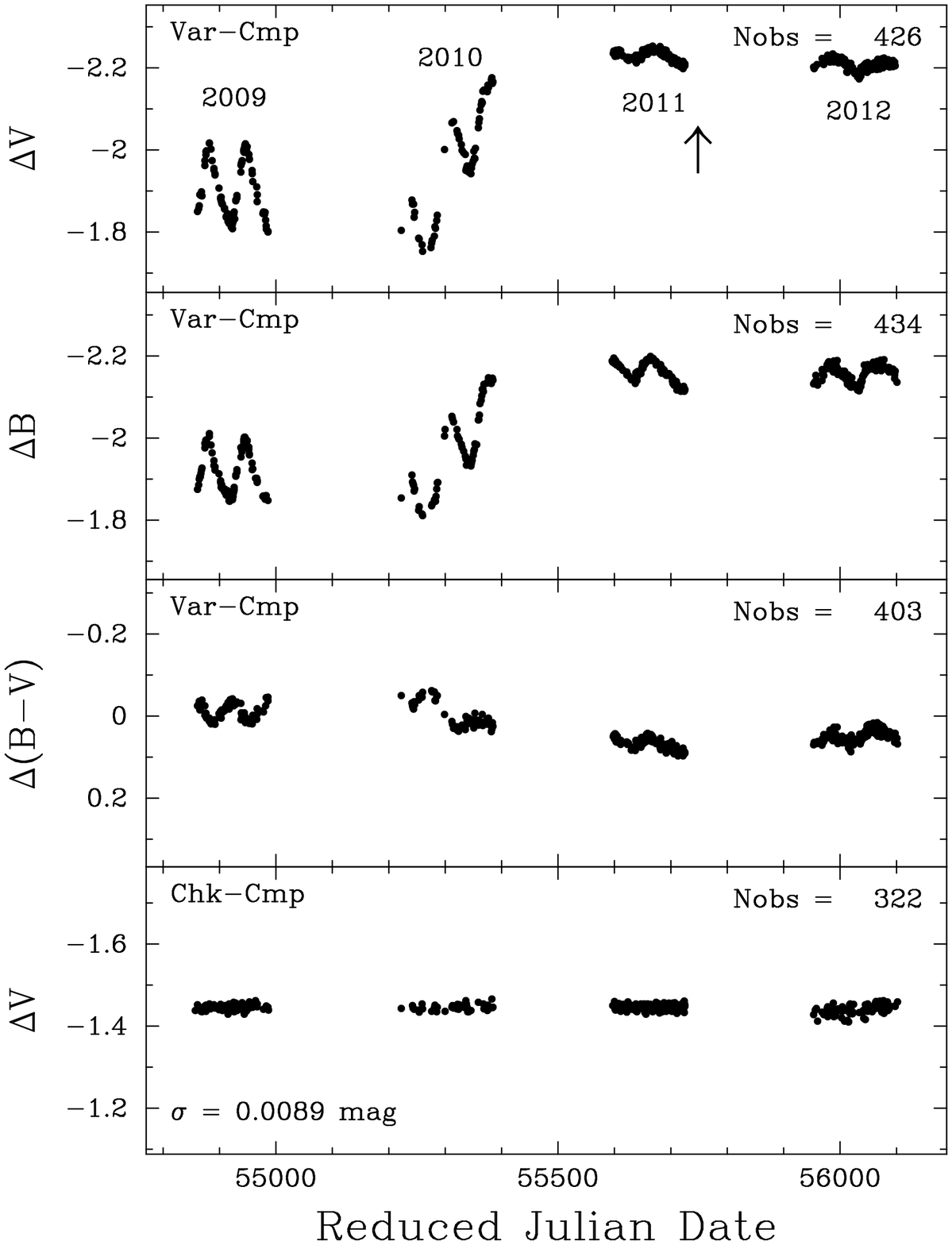}
\caption{Photometric data of $\delta$~Sco acquired over four years with the
T3 0.4~m APT at Fairborn Observatory. The vertical and horizontal scales for all four panels
are identical to facilitate direct comparison of the brightness variations 
in the $V$ and $B$ filters, in the $B-V$ color index, and brightness changes 
in the K$-$C differential magnitudes.  The arrow in the top panel marks the
time of the latest periastron passage on UT 2011 ${\rm July}~6 \pm 2$ days. The total number of observations is given in the upper right of each panel. The standard deviation, $\sigma$, of the $K-C$ is provided in the lower left of the figure.}
\label{photo_all}
\end{figure}

As previously mentioned, $\delta$ Sco is highly variable and this behaviour can clearly be seen in Figure~\ref{photo_all}. The arrow in the top panel of this Figure marks the periastron passage of the secondary and reveals no significant change in the photometry surrounding this time. Cyclical behavior with timescales of approximately 60--100 days is visible in all four seasons with obvious differences in amplitude and mean brightness of the system. This variability of similar frequency was observed much earlier by \citet{gan02} so these variations have persisted for at least a decade, even at times when the disk density was much different.  As mentioned previously, the $V$-band photometry is sensitive to the emitting volume within the first few stellar radii (see figure 1 of \citet{car11}). Therefore, this variability could be due to changes in the density of the disk within the first few stellar radii or as a result of shielding of the inner disk by warping or flaring of the outer disk.  

The top two panels of Figure~\ref{photo_09_10} show that the brightness and color 
index in 2009 were anti-correlated, i.e., the star became redder as it got 
brighter.  These trends are relatively straight forward to explain.  As the system brightens in the $V$-band with the injection of additional gas into the disk, the system becomes correspondingly redder due to an increase in radiative processes, especially free-free emission that increases with wavelength. The bottom two panels demonstrate the same relation in 2010.  
However, the top two panels of Figure~\ref{photo_11_12} show a reversal in the sense of
the correlation; the star gets bluer as it gets brighter.  Finally, the 
brightness and color changes for 2012, plotted in the bottom two panels 
of Figure~\ref{photo_11_12}, appear to be significantly out of phase.

We determined individual photometric periods for all four observing seasons in both the $B$ and $V$ passbands. The complete results are given in Table~\ref{periods}. Figure~\ref{plot_periods} shows one example of our period determination, for the $B$ data of 2010. The frequency spectrum is computed by fitting least-squares sinusoids to the data over a range of trial frequencies and finding the frequency that best reduces the total variance of the data set.  Because of the slope in the observations for 2010 and 2011, we first removed a linear trend from the data before computing the frequency spectrum and the corresponding phase curves. 

Table~\ref{periods} shows that both the photometric period and amplitude of $\delta$~Sco varied dramatically from year to year, from 64 to 94 days and 0.03 to 0.18 mag, respectively. One result, namely the 2012 $V$ period determination, may be suspect. The 2012 $V$ light curve has a much lower S/N ratio than most of the 8 light curves. In addition, the comp 
star exhibits possible low-amplitude variability for the first time near the end of the 2012 observing season.  This combination renders the period determination for the 2012 $V$ light curve unreliable.

\begin{deluxetable}{ccccc}
\tablenum{2}
\tabletypesize{\normalsize}
\tablewidth{0pt}
\tablecaption{YEARLY PHOTOMETRIC PERIODS FOR $\delta$~SCO \label{periods}}
\tablehead{
\colhead{Observing} & \colhead{Photometric} & \colhead{$N_{obs}$} & 
\colhead{Period} & \colhead{Full Amplitude} \\
\colhead{Season} & \colhead{Passband} & \colhead{} & \colhead{(days)} & 
\colhead{(mag)} \\
\colhead{(1)} & \colhead{(2)} & \colhead{(3)} & \colhead{(4)} & \colhead{(5)} 
}
\startdata
 2009 & $B$ &  91 &  $64.3\pm2.4$ & $0.141\pm0.004$  \\
      & $V$ &  94 &  $64.9\pm2.5$ & $0.179\pm0.005$  \\
 2010 & $B$ &  64 &  $72.1\pm2.5$\tablenotemark{a} & $0.156\pm0.005$  \\
      & $V$ &  61 &  $73.0\pm3.0$\tablenotemark{a} & $0.166\pm0.009$  \\
 2011 & $B$ & 127 &  $94.3\pm3.6$\tablenotemark{a} & $0.053\pm0.002$  \\
      & $V$ & 126 &  $94.2\pm3.5$\tablenotemark{a} & $0.031\pm0.002$  \\
 2012 & $B$ & 152 &  $83.1\pm2.5$ & $0.047\pm0.002$  \\
      & $V$ & 145 & $101.6\pm4.5$\tablenotemark{b} & $0.027\pm0.002$  \\
\enddata
\tablenotetext{a}{Due to the slope in these light curves, we first removed
a linear trend before applying the period analysis.}
\tablenotetext{b}{This period determination is probably much more uncertain
than the formal error indicates.  See the text for details.} 
\end{deluxetable}

Figure~\ref{all} shows the $\Delta V$ magnitude versus $\Delta(B-V)$ for all four observing seasons with each season  2009, 2010, 2011, 2012 indicated by the four shades of gray from light to dark, respectively. The general trend of the data from the bottom left to upper right is due to the  increase in V magnitude and a reddening of colour during the four years of observations as the disk builds. Note that the $\Delta(B-V)$ colour index becomes redder (more positive) as expected with the building of the disk. Overall the $V$ magnitude begins to saturate and then most variations are changes mainly in colour. This is particularly evident in the observations for 2011 and 2012 since the changes in $V$ magnitude are not as significant compared with 2009 and 2010. Also note that the 2012 season is not quite as bright as the 2011 season.  It will be interesting to see if this trend continues in future years.

It is also interesting to replot Figure~\ref{all} as a function of time for each observing season.  We start with Figures~\ref{loop1} and \ref{loop2} showing $\Delta V$ versus $\Delta(B-V)$ for the 2009 season as a function of time. Although $\delta$ Sco is not as bright in the $V$-band in 2009, this season shows two well-defined increases in the $V$- and $B$-band,  both of which return to near their original values at the start of the season (see Figure~\ref{photo_all}). Figure~\ref{loop1} covers the period starting from the Reduced Julian date 54861 for 66 days, and Figure~\ref{loop2} covers the remaining season of 76 days beginning at Reduced Julian date 54928. The division of the 2009 observing season into these two parts was rather arbitrary; we tried to divide the data where the loop seemed to repeat. Each figure shows a distinct loop during the specified time-frame due to an increase in brightness and reddening that is followed by a subsequent decline in brightness and by a slow return to bluer color-index. Notice in Figures~\ref{loop1} and \ref{loop2} that the maximum and minimum bounds of $\Delta V$ and $\Delta(B-V)$ are remarkably similar. This is not surprising since Figure~\ref{photo_all} shows that the peak brightness for both $\Delta V$ and $\Delta B$ are almost identical in 2009. Figure~\ref{season2} shows that, in 2010, we also have two loops. However, the system remains bright in the $V$-band after the first loop, so the second loop starts at brighter values of the $V$-band and continues to increase during the formation of the loop. This can also be seen in Figure~\ref{photo_all} for the 2010 season.  At the end of the 2010 observing season, both the $V$- and $B$-band magnitudes remain well above their values at the beginning of this season. \citet{dew06} studied the light and colour variability of Be stars in the Small Magellanic Cloud and found similar loops to those displayed in Figure~\ref{loop1} to Figure~\ref{season2} for 40\% of the stars having a photometric variability of greater than 0.2$^{\rm m}$. Among their group of stars that exhibited this behaviour, 90\% of these loops were traced out in a clockwise direction analogous to the behaviour we find for $\delta$ Sco. \citet{dew06} attributed clockwise loops to slowly outward flowing material as a disk builds followed by an emptying of the disk from the inside-out as mass loss from the stellar surface is terminated. They suggest that anticlockwise loops are indicative of accretion in young systems such as the Herbig Be stars. 

In 2011 and 2012, the loops are not as distinctive since the changes in  $\Delta V$ and $\Delta B$ are smaller (see Figure~\ref{photo_all} and Figure~\ref{all}) so we have not included plots of these two observing seasons.  Nevertheless, as mentioned previously, the cyclical variability is still noticeable (see Figure~\ref{photo_all}). A movie showing the $\Delta V$ versus $\Delta(B-V)$ as a function of time for all four seasons is available in the online version of the paper.
 
We also produced models of the $\delta$ Sco system for the recent periastron passage following the approach by \citet{hol99}. These models rely on gravity to follow the particles orbiting in a Keplerian fashion within the disk; gas dynamics are not included. We adopt the star/disk system and orbital parameters from \citet{car06sco} and \citet{tyc11}, respectively. The total mass of the disk for our simulation of $3 \times 10^{-8}$  M$_\odot$ was calculated from the base density, $\rho_o = 4.5 \times 10^{-10}$ g cm${^{-3}}$ \citep{car06sco} with an average power-law density fall off of 3.5 for a disk truncated at 7 R$_{*}$. While the mass of $\delta$ Sco's disk certainly changes over time, and perhaps substantially in the inner disk during mass ejection, this is a reasonable estimate of the total disk mass for our preliminary modeling. Figure~\ref{tide} shows the system 3.4 years after periastron passage. For this particular simulation, we assume that the disk, the equatorial plane of the star and the secondary all lie in the same plane.  Although the secondary does not directly impact the disk, material is pulled away from the disk, forming a tidal tail that extends out to 5 AU at the time of the snapshot. Assuming a distance of 135 pc \citep{tyc11}, the tail size is equivalent to $\sim$40 milliarcsesconds. Over time, the tail will continue to expand. Despite the fact that the secondary does not impact the disk, it does change the geometry of the disk system and these changes could potentially block parts of the primary star and inner disk from the field of view where the largest contribution to the $V$-band is produced. However, since the orbital time scale for the secondary passage is of order of a decade, there must be some other process that is responsible for the cyclical variability. We note, however, that if the disk size is much smaller than 7 R$_{*}$, then a tidal tail will not form since disks smaller than 
5 R$_{*}$ do not produce tails in our simulations.  We also note that tail production is curtailed if the secondary orbit is not in the same plane as that of the disk. This prediction is consistent with \citet{che12}, who find that the secondary passage did not cause any mass outflow during periastron.

\section{Discussion and Summary}
\label{sum}

Nonradial pulsations are generally believed to be the source of short term variability, especially for the early-type Be stars where they are observed to be present in 80 - 90\% of the disk systems \citep{por03}. Typical periods vary from less than a day to several days \citep{riv03}.
Time scales corresponding to the viscous disk model (discussed above) are much longer, from $10^2$ to $10^3$ days \citep{jon08}. The cyclical variability observed in $\delta$ Sco is intermediate in length and it is interesting to speculate on the possible mechanism required to produce variability on such a scale. 

A possible explanation for the variation in $V$-band magnitude is that $\delta$ Sco is subjected to both gravity darkening caused by rapid rotation and precession due to an unseen third companion. If so, as the star precesses, different latitudes would come into view. Because the star is gravity darkened the variation in the effective temperature from pole to equator causes the amount of light radiated to change with latitude. The star's apparent brightness would therefore change as the star is observed at different times during its precession.

To test the effect of precession, a gravity darkened star was simulated and the difference in apparent flux in the visual band was estimated for different inclinations of the star. To simplify the calculation, the star was assumed to be emitting as a perfect black body. The star was divided into several latitudinal sections, each of which are considered to be at constant effective temperature. Then, for the considered inclination angle, the projected area of each latitudinal section was calculated. For simplicity, the star was assumed to be a perfect sphere instead of an oblate spheroid. To estimate the relative apparent flux at a specific inclination angle, the projected area and effective temperature of each latitudinal section was calculated. The effective temperature was deduced by calculating the average surface gravity for each latitudinal section and using the von Zeipel theorem (that states that the effective temperature is proportional to fourth root of the local surface gravity \citep{von24}). The total relative apparent flux in the $V$-band for a particular inclination angle was then obtained by adding up the relative apparent flux of each section and integrating over the $V$-band.

Two cases were considered. The first case was for a precessing star without any obstruction by a disk, while in the second case, a static disk was added. It was assumed that the structure of the disk was unchanging and that it completely blocks the $V$-band emission of the obstructed part of the star. The results of these tests showed that even for the most extreme case by allowing the inclination to vary pole-on to edge-on, the changes in $V$-band magnitude were orders of magnitude lower than the observed variation. Therefore it is unlikely that the variation in magnitude can be attributed to a precession of the star. We note that the critical velocity of $\delta$ Sco is 620 km/s and with an inclination of $38\pm5^{o}$ \citep{car06sco} combined with observed vsini of 148 km/s \citep{bro97} means that $\delta$ Sco is rotating well below critical by approximately 40\%. For more rapidly rotating Be stars, the variation in temperature from pole to equator would be more significant so that it would be possible for precession to have a larger effect in other stars.

\citet{car06sco} considered whether or not a change in disk geometry could help explain the optical fadings observed for $\delta$ Sco. They concluded, based on $\delta$ Sco's low inclination, that a significant warping of the disk could block enough of the stellar surface to account for the fadings if enough material was moved to higher latitudes. In fact, tidal warping is frequently invoked to explain Be star transitions from singly or doubly peaked emission lines to shell lines or vice versa \citep{mar11}. The size of the H$\alpha$ emitting region should be roughly equivalent to the size of the disk \citep{mir03} with the optical fading anti-correlated with H${\alpha}$ line strength (see \citet{mir11,car06sco}). Recall, as discussed above, the $V$-band contribution will originate in a volume of gas very close to the star and not over the full extent of the H$\alpha$ emitting region. Since $\delta$ Sco's H$\alpha$ line transitioned from doubly to singly peaked in 2003 \citep{mir03}, there is evidence to support this suggestion. The fact that the fadings have been observed to be anti-correlated with line emission \citep{mir03} seems to also require an increase in disk mass. Interestingly, \citet{smi12} noted for the binary system $\gamma$ Cas that the quasi-secular brightening occurred in the optical during 2010, and early 2011 was correlated with a higher column
density which was manifested by an attenuation of the soft X-ray flux. Figure~\ref{tide} shows how material could be stripped from the disk at periastron, which could potentially shield portions of the inner disk where the disk $V$-band is produced. However, the $\sim$10 year binary period is much too long to explain the cyclical variability.

It is also interesting to consider whether or not an unseen companion could cause the periodic variations. If a yet unseen companion orbits the primary with a 70 day period, for example, we can determine the semi-major axis, a, of its orbit from Kepler's Third Law. Assuming a stellar mass of 12.4M$_{\odot}$ for the primary \citep{tyc11}, we find $\rm{a} = 0.77$ AU. Here we assume that the mass of the unseen companion is much less than the mass of the primary. \citet{tyc11} give a radius for the primary of 0.45 milliarcsecs.
If $\delta$ Sco is at a distance of 135 pc from Earth, the primary radius is
then 0.061 AU, and a is 0.77/0.061 = 13R${_{*}}$. Thus, if an unseen companion orbiting the primary is the source of the cyclical variability (70 days in this example), it must orbit the primary near this distance. The minimum separation at periastron of the secondary is 14 R${_{*}}$ \citep{tyc11}. This is very near the size of the orbit of the putative unseen companion and raises the question of whether or not the companion could remain in such an orbit in the long-term or would be destabilized by repeated passages of the secondary. \citet{hol99} examined the stability of planets in binary star systems, and we can use their result to examine the stability of the hypothetical unseen companion.

Taking the secondary mass to be 8 M$_{\odot}$ \citep{tyc11}, we can use
their expression (1) to calculate that the smallest stable orbit
expected around the primary is less than the primary star's radius. We note, however, that $\delta$ Sco has a more extreme eccentricity, $e = 0.938$, \citep{tyc11}
than examined by \citet{hol99}, who considered a maximum of 0.9, so
our calculation above takes their equation outside its regime of
applicability. If instead we use e = 0.9 (their maximum value) we get a
stability limit of 2.8 R$_{*}$; thus, our hypothetical companion on
a 70 day orbit is still outside the regime of stability under this
relaxed assumption. This does not mean that such a companion could not
survive a single periastron passage at 13 R$_{*}$, only that it
cannot do so over many such passages. Thus, it seems unlikely that an
unseen companion with a $\sim$ 70 day orbital period around the primary is the
cause of the observed variation, as such an orbit is unstable.

We are currently modeling the periodic loops in the $\Delta V$ versus $\Delta(B-V)$
colour-magnitude diagram with sophisticated computational codes. Our
initial models simulate the addition and depletion of material in the
inner region of an axisymmetric circumstellar disk constructed using a
conventional power-law radial distribution. Gas is added to the inner
edge of the disk during predetermined periods of constant mass loss.
Similarly, gas is removed from the inner edge of the disk once mass
transfer from the star to the disk ends. The thermal structure of the
disk is recomputed after each adjustment to the distribution of gas in
the disk using the radiative transfer code of \citet{sig07}.
The theoretical observables are calculated using the Monte Carlo
simulation of \citet{hal12}. Using this procedure, we can
analyze the evolving physical conditions of the circumstellar gas
through trends in predicted observables such as the colour-magnitude
diagrams.

\begin{figure}
\epsscale{0.75}
\plotone{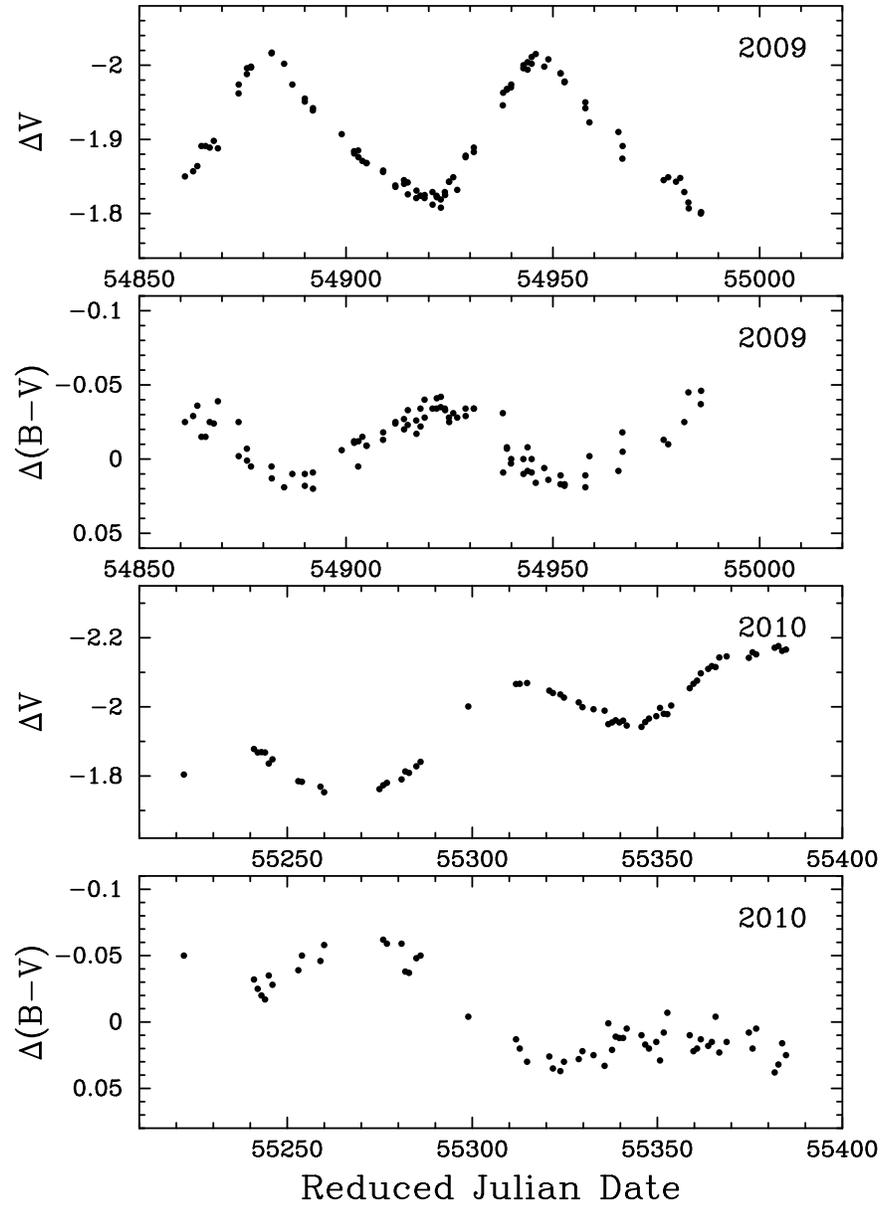}
\caption{Top two panels: Individual $\Delta V$ magnitudes and
$\Delta(B-V)$ color indices for 2009.  Brightness and color are
anti-correlated.  Bottom two panels:  Same as the top two panels but for 2010.}
\label{photo_09_10}
\end{figure}

\begin{figure}
\epsscale{0.75}
\plotone{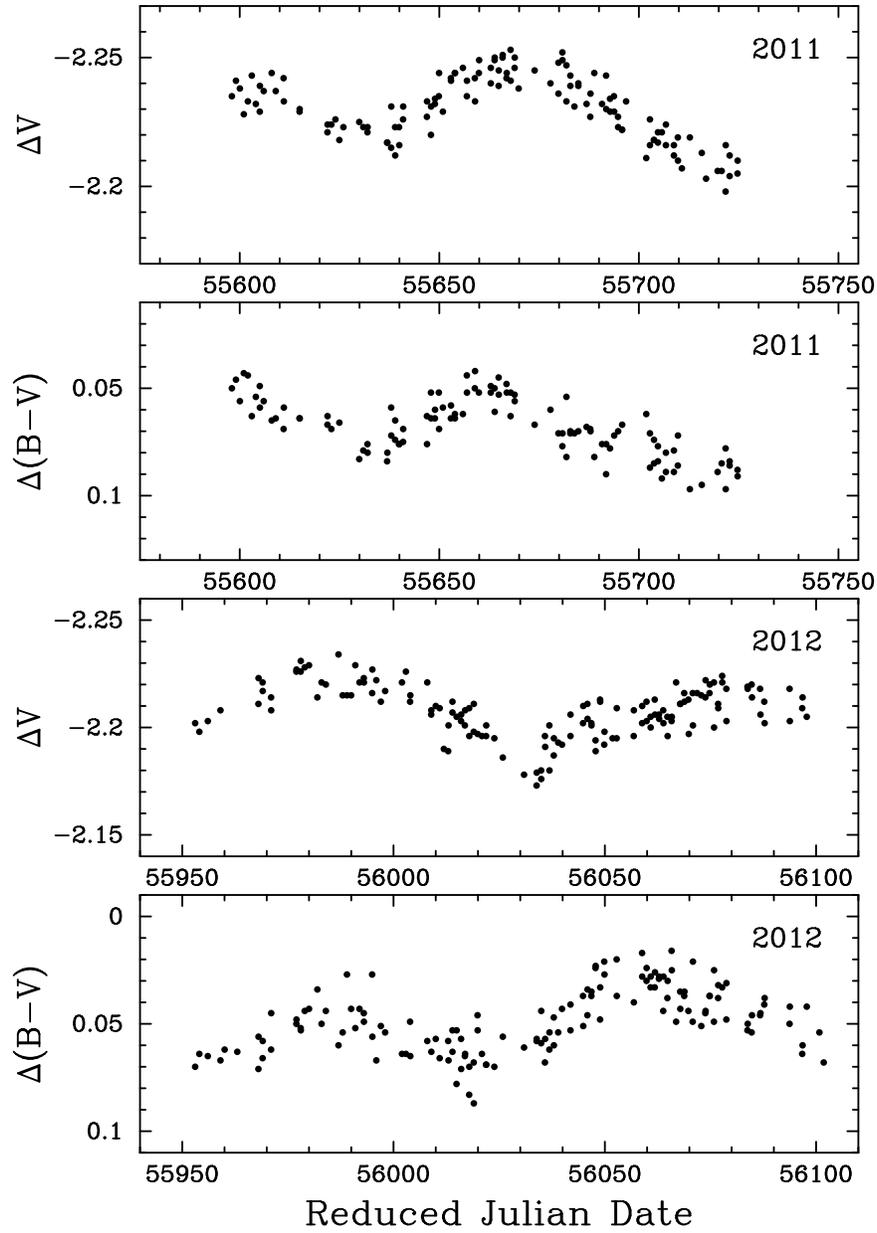}
\caption{ Top two panels: $\Delta V$ magnitudes and $\Delta(B-V)$ color
indices for 2011.  Brightness and color are now directly correlated.
Bottom two panels:  Same as the top two panels but for 2012. $V$ magnitudes and $(B-V)$ color indices appear to
be significantly out of phase.}
\label{photo_11_12}
\end{figure}

\begin{figure}
\epsscale{1.0}
\plotone{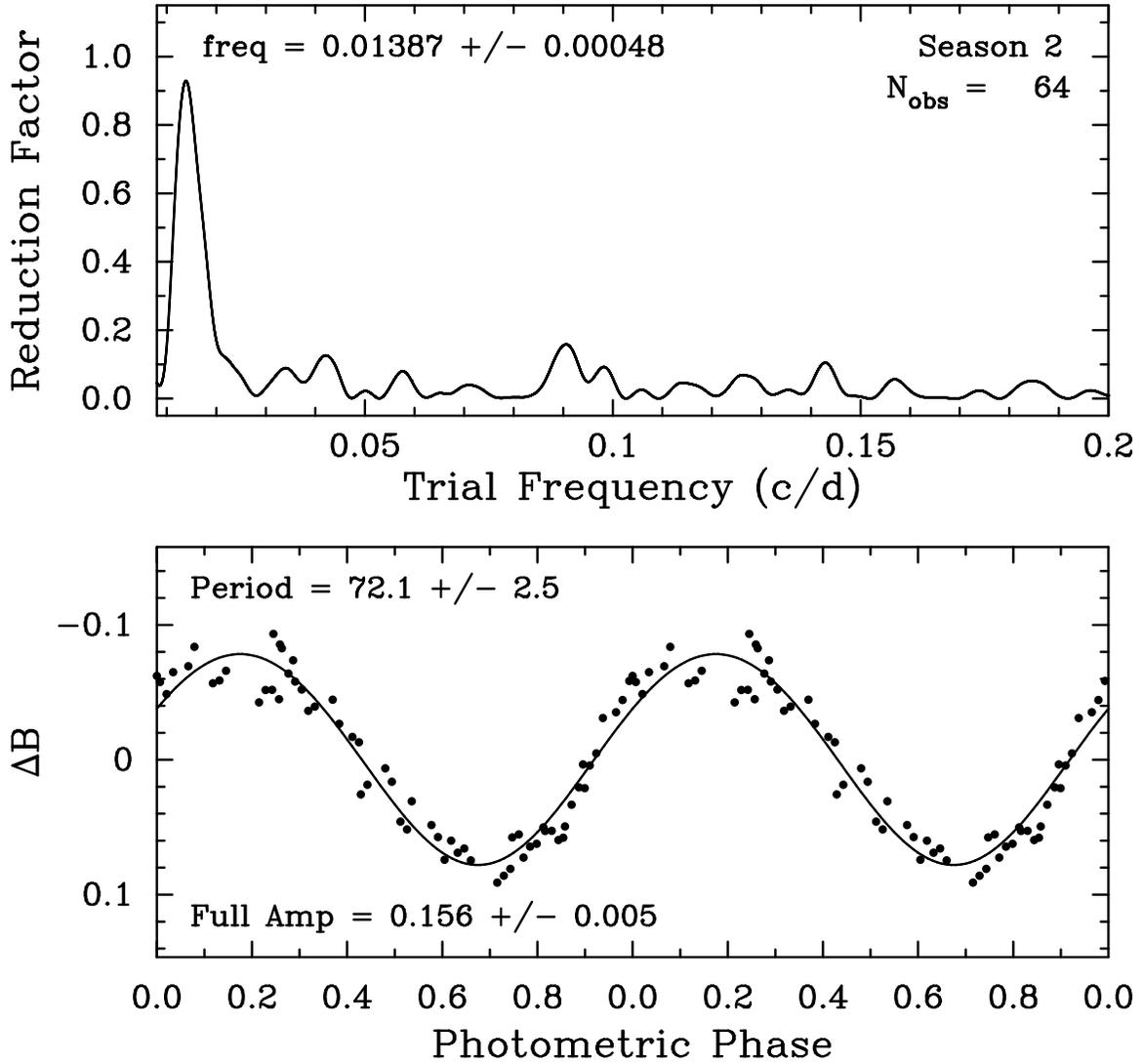}
\caption{$Top$: Frequency spectrum of the 2010 Johnson $B$ photometry after removing a linear trend from the data.  Best frequency is 0.01387 c~d$^{-1}$. $Bottom$: $B$ data from 2010 phased with the corresponding best period of 72.1 days.  The peak-to-peak amplitude is 0.156 mag.  See Table~2 for the complete results of our period determinations.}
\label{plot_periods}
\end{figure}

\begin{figure}
\plotone{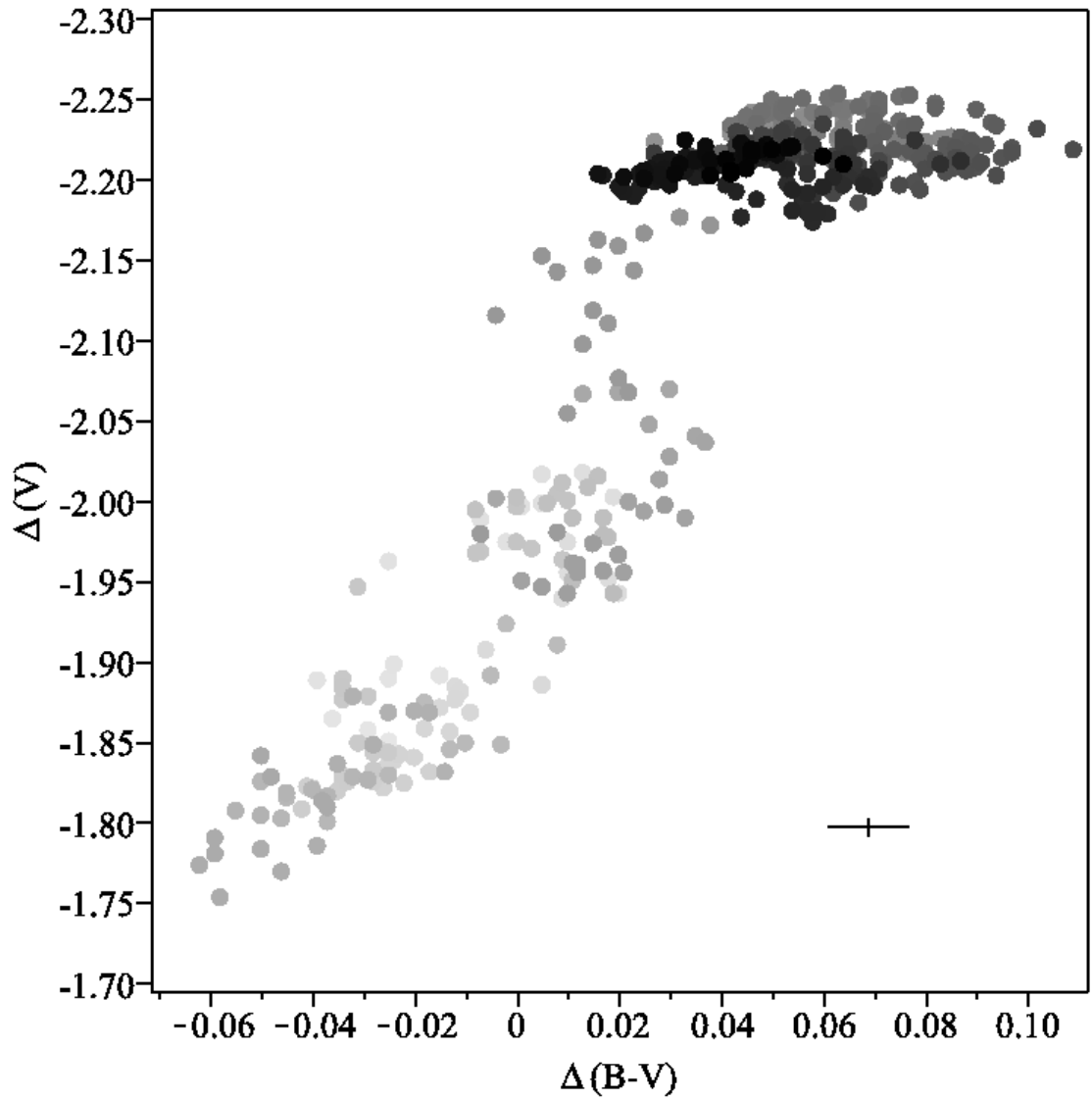}
\caption{This figure shows the $\Delta V$ magnitude versus $\Delta(B-V)$ for all four observing seasons. The 2009, 2010, 2011, and 2012 seasons are indicated by the four shades of gray dots from light to dark, respectively. The error estimate for each value is represented by one error bar in the lower right of the figure.}
\label{all}
\end{figure}

\begin{figure}
\plotone{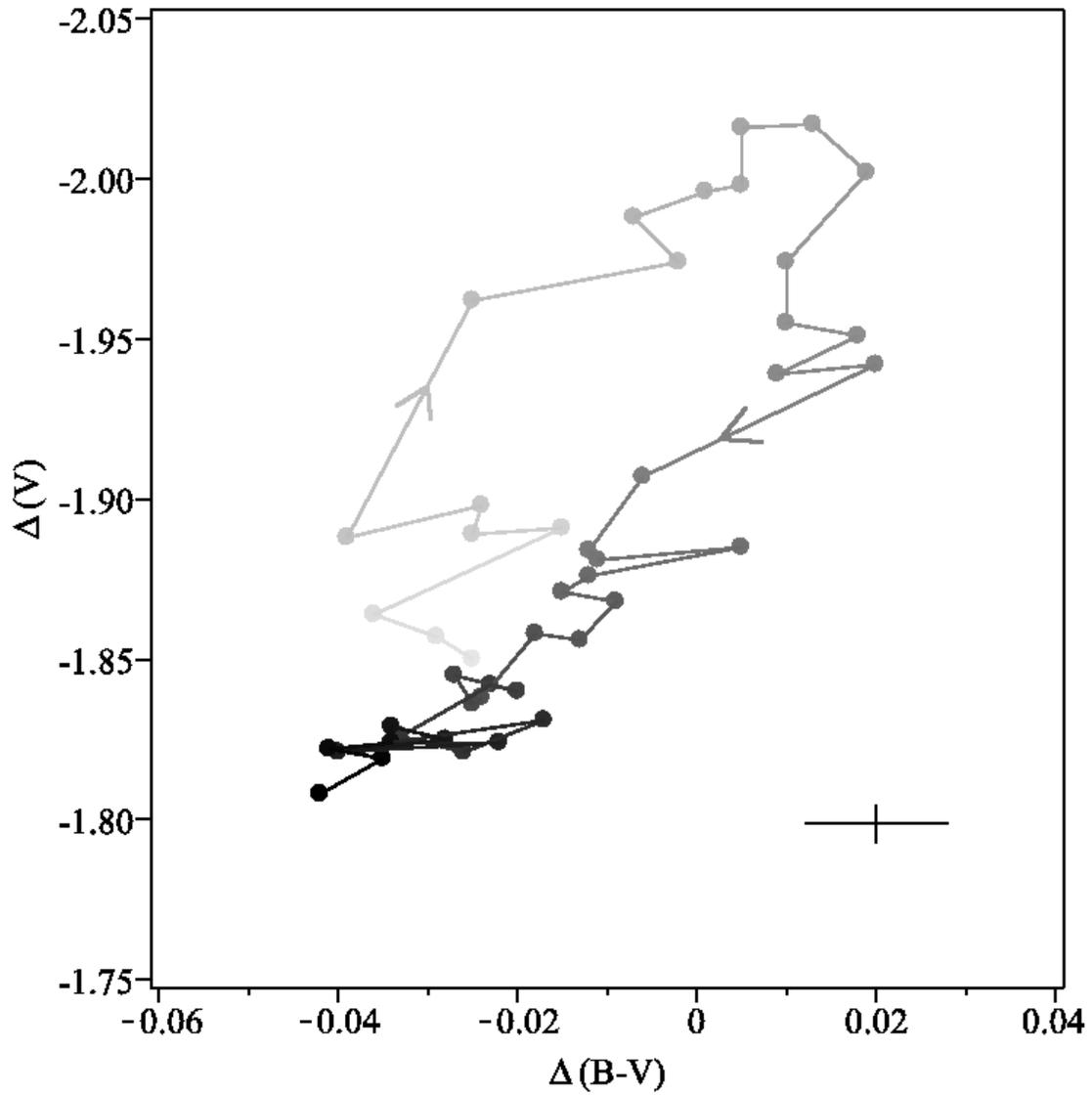}
\caption{This figure shows $\Delta V$ versus $\Delta(B-V)$ as a function of time for the first 65 days of the 2009 observing season from the Reduced Julian dates 54861-54922. The color from light gray to dark gray indicates the progression with time. The observational error for each data value is given in the lower right. The lines connect the data in time with the arrows indicating the sense of direction.}
\label{loop1}
\end{figure}

\begin{figure}
\plotone{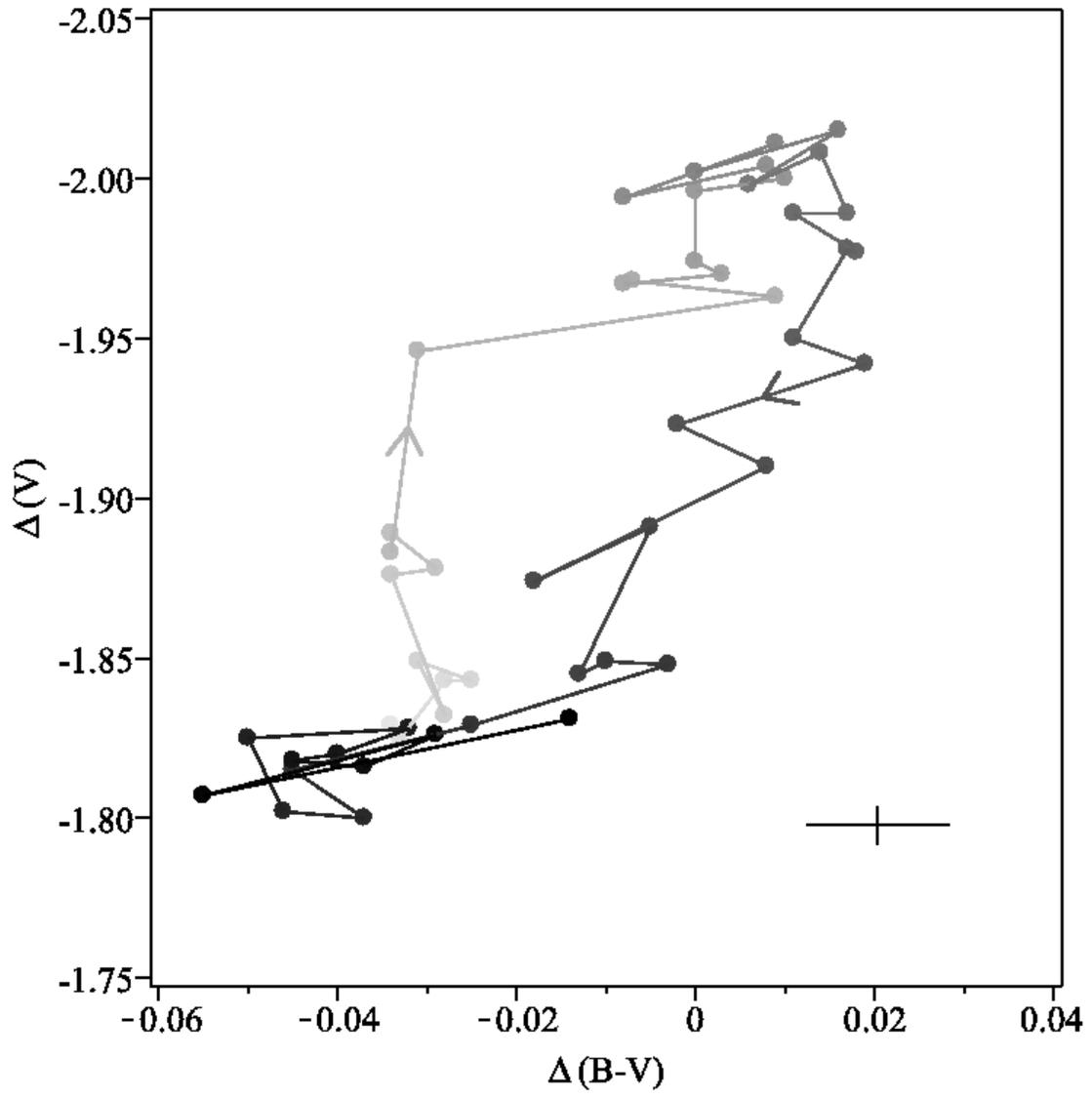}
\caption{Same as Figure~\ref{loop1} for the second half of the observing season corresponding to Reduced Julian dates 54923-55006.}
\label{loop2}
\end{figure}

\begin{figure}
\plotone{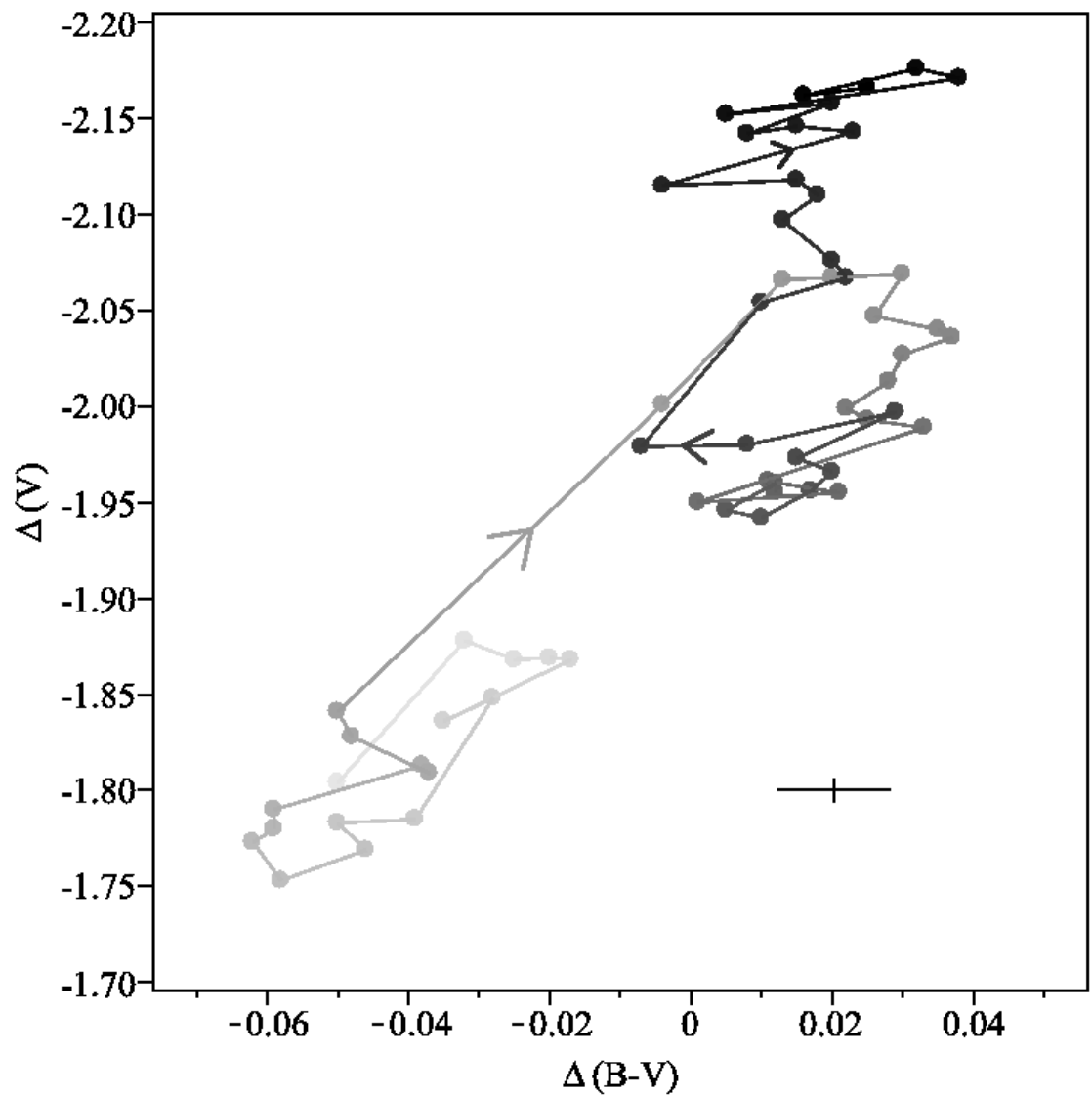}
\caption{Same as Figure~\ref{loop1} for the entire 2010 from Reduced Julian dates 55222-55384.}
\label{season2}
\end{figure}

\begin{figure}
\plotone{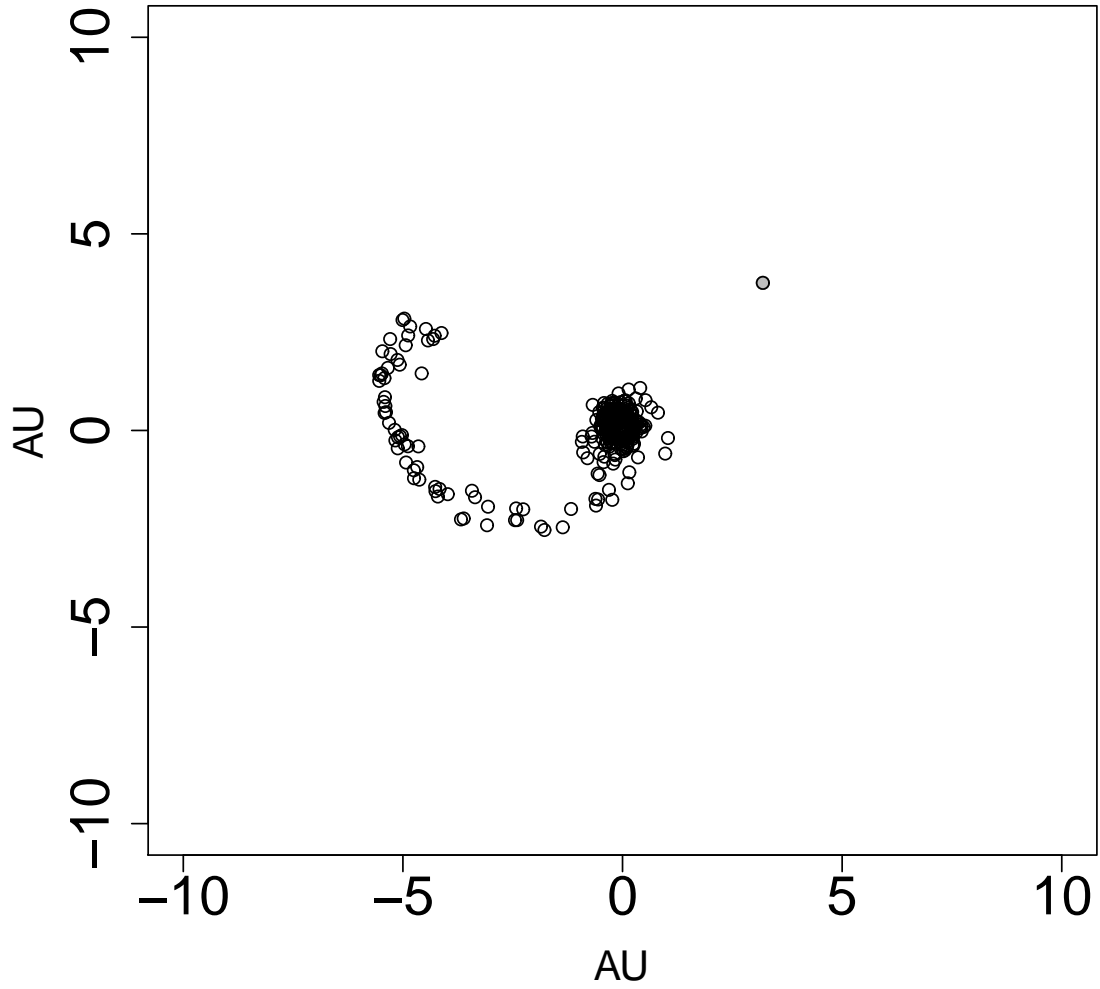}
\caption{Disk particles (open circles) after periastron passage of the secondary (shown by the grey dot in the upper right). In this case, the disk is 7 R$_{*}$ in radius so the secondary does not impact the disk or capture any material from the system. Instead, a dramatic tidal tail is produced.}
\label{tide}
\end{figure}

Our results indicate that we can reproduce the observed
changes in the magnitude and colour of the star using models with
different initial parameters. By changing the base density of the disk
and the size of the inner region that is cleared and refilled, we have
determined that the short-term mass-loss rates that occur over roughly
35 days range from $2.5 - 7.0 \times 10^{-9}$ $M_{\sun}/$yr. These models reproduce the
photometric trends plotted in Figure~\ref{photo_09_10} that show the system becoming
redder as it gets brighter. We expect that assiduous comparison of the
shapes of the observed and predicted loops will reduce much of the
degeneracy between the models and further constrain our predicted
mass-loss rates. While we have initially restricted ourselves to a
disk truncated at 7 R${_*}$ to be consistent with \citet{car06sco},
we acknowledge that the truncation radius of the disk is an additional
parameter that should be constrained by other observations. \citet{riv12} suggest that, after the periastron in 2011, a disturbance propagated inward throughout the disk. It is interesting to speculate that perhaps this caused the disk to become smaller in 2011 and 2012. We note that the photometric cycles in the $B$ and $V$ passbands vary in length and amplitude (see Table~\ref{periods}) from year to year. This interesting complication will also require further study. It is
clear that detailed modeling, including changes in disk geometry,
episodic and/or asymmetrical mass-loss rates must be invoked to
explain all of the observed peculiarities of this system. We are
currently constructing models with various densities, disk sizes,
evacuated regions and other important disk properties to see if the
observations can be used to constrain the nature of the changing
condition in the disk. The presentation and analysis of these models
is the focus of a follow-up paper.

\acknowledgements C.E.J. and R.J. H. acknowledges research supported by NSERC, the Natural Sciences and Engineering Research Council of Canada. Astronomy at Tennessee State University is supported by NASA, NSF, Tennessee State University, and the state of Tennessee through its Centers of 
Excellence programs.

\end{document}